\documentstyle[amsfonts,preprint,prl,aps,12pt]{revtex}

\begin{document}


\title{\bf Annihilation catastrophe: new critical phenomenon in
diffusion-controlled dynamics}
\author{ Boris ~M.~Shipilevsky}
\address{ Institute of Solid State Physics, Chernogolovka,
Moscow district, 142432, Russia}
\maketitle



\baselineskip=14pt
\vspace*{1cm}

 For the last decade the reaction-diffusion system $A + B \rightarrow 0$,
where unlike species $A$ and $B$ diffuse and irreversibly react in the bulk
of a $d$-dimensional substrate, has acquired the status of one of the most
popular objects in nonequilibrium statistical physics [1].
Two problems have been investigated most intensively: (i) dynamics of the
$A +B \rightarrow 0$ annihilation in an infinite system with initially
homogeneously (randomly) and equimolarly distributed reactants, in which
below the critical dimension $d_{c}=4$ a dynamical clustering develops
(fluctuation-induced like-particle domains formation) and, as a result,
an anomalous reaction deceleration arises (Ovchinnikov and Zeldovich,
1978 [2], Toussaint and Wilczek, 1983 [3]); (ii) behaviour of the
dynamic reaction front in an infinite system with initially spatially
separated reactants (Galfi and Racz, 1988 [4]), and structure of the
steady state front in a finite system, at the ends of which are injected
equal currents of $A's$ and $B's$ particles
(Ben-Naim and Redner, 1992 [5], Cornell and Droz, 1993 [6]).

In 1994 it was discovered (Shipilevsky, 1994 [7]) that in another
wide class of RD systems, where reaction and diffusion are spatially
separated (i.e. reaction proceeds on the surface of the medium and diffusion
proceeds in its bulk) the interplay between reaction and diffusion acquires
qualitatively new features and leads to the threshold self-organizing 
dynamics of the $A+B\rightarrow 0$. It has been found that once particles $A$ 
and $B$ diffuse at different mobilities from the bulk of finite medium onto 
the surface and die on it by the reaction $A + B \rightarrow 0$, there should 
exist some threshold difference in the initial numbers of $A$ and $B$ 
particles, $\Delta_{c}$, above which the loop of positive feedback is 
"switched on" and the process of their death, instead of usual deceleration, 
starts {\it to accelerate autocatalytically}. Recently, I have shown [8] 
that deceleration-acceleration transition is only a prelude to much more
nontrivial dynamical effects, which take place far beyond the acceleration
threshold. It has been demonstrated that in the diffusion-controlled limit 
$\Delta \rightarrow \infty$ a new critical phenomenon develops, which arises 
as a result of self-organizing explosive growth (drop) of the surface 
concentrations of, respectively, slow and fast particles ({\it concentration 
explosion}) and manifests itself in the form of an abrupt singular jump in 
the desorption flux relaxation rate. The aim of this report is to give an 
insight into the key features of this remarkable phenomenon which I call 
{\it annihilation catastrophe}.

{\bf Model.} I consider a model, in which species $A$ and $B$ are supposed to
be initially uniformly distributed in the bulk of infinitely extended slab of 
thickness $2\ell$. Both species diffuse to the surface $(X=\pm\ell)$ and 
desorb as a result of surface reaction 
$A_{ads}+B_{ads}\rightarrow AB \rightarrow 0$. 
The rate of reaction is proportional to the product of surface concentrations
$I=k\rho_{A}\rho_{B}=\kappa c_{As}c_{Bs}$, where the rates of
surface-subsurface exchange are supposed to be rather high, so that
quasiequilibrium $\rho_{i}=f_{i}c_{is}$ is always sustained. The system is
effectively one dimensional and the boundary conditions are determined from
the equality of diffusion $I^{D}$ and desorption $I$ flux densities at the
surface $I^{D}\mid_{s}=I$, i.e. it is assumed that the surface layer
capacity can be neglected. Thus, introducing index $"H"$({\it heavy}) for
slower diffusing species and index $"L"$({\it light}) for a faster one,
the problem reads ( by symmetry I consider the interval [$0,\ell$] only)

\begin{eqnarray}
\partial h/\partial \tau  = \nabla^{2}h \quad , \quad
\partial l/\partial \tau  = (1/p)\nabla^{2}l,
\end{eqnarray}
\begin{eqnarray}
\nabla h \mid_{s} = (1/p)\nabla l\mid_{s}
= - h_{s}l_{s},
\end{eqnarray}

with $\nabla (h,l) \mid_{x=0}=0$ and the initial
conditions $h(x,0)=h_{0}$ and $l(x,0)=l_{0}$. Here $h(x,\tau)=c_{H}/c_{*}$
and $l(x,\tau)=c_{L}/c_{*}$ are the reduced concentrations,
$\nabla\equiv\partial/\partial x, x=X/\ell\in [0,1]$ is the dimensionless
coordinate, $\tau= D_{H}t/\ell^{2}$ is the dimensionless time,
$p=D_{H}/D_{L} \leq 1$ is the ratio of species diffusivities, and 
$c_{*}=D_{H}/\kappa\ell$ is the characteristic concentration scale, from
which diffusion starts to play an essential role. Boundary conditions
(2) reflect the fact that particles disappear in pairs only, i.e.
$J=h_{s}l_{s}= -<\dot{h}>=-<\dot{l}>$, where $J=I/I_{*}$ is the reduced
desorption flux density and $I_{*}=\kappa c_{*}^{2}$ is its characteristic
scale, therefore $<h> - <l> = \Delta = {\rm const.}$, i.e., the excess amount 
stays "inert" in the bulk ( here $<h>=\int_{0}^{1}h dx=
{\cal N}_{H}/{\cal N}_{*}$ and $<l>=\int_{0}^{1}l dx =
{\cal N}_{L}/{\cal N}_{*}$ are the total reduced numbers of particles in the
bulk per unit of surface and ${\cal N}_{*}=c_{*}\ell=D_{H}/\kappa$ is the
characteristic particle number scale). I consider here the annihilation 
dynamics for $\Delta > 0$ when the pair number $N$ is dictated by the $L$ 
particles number $N(\tau)={\cal N}_{pair}/{\cal N}_{*}= N_{L}(\tau)$ so 
that in the final state all $L$ particles disappear 
$N(\infty)=l(x,\infty)=0$ and $H$ particles are distributed uniformly with 
concentration $h(x,\infty)=\Delta$. I am interested first and foremost in 
the diffusion-controlled limit $N_{0}\rightarrow \infty$ (note that 
diffusion-controlled limit $N_{0}=l_{0}=c_{L}(0)/c_{*}\to\infty$ means 
$c^{-1}_{*}=\kappa\ell/D_{H}\to\infty$). Moreover, I focus on the limit 
$n_{0}=N_{0}/\Delta \gg 1$ when the initial numbers of $H$ and $L$ particles 
are almost equal, $r=h_{0}/l_{0}=(1+n_{0})/n_{0}\approx 1$. Precisely in 
this limit the particles interplay is most "intensive" and gives rise to 
the explosive dynamics, that is my prime interest here.    

{\bf Transition point and self-accelerating annihilation.}
Assuming $N_{0}\rightarrow \infty$ and ${\cal R}=r\sqrt{p} < 1$, I show [10],
that for $n_{0} \gg 1$ and $\gamma= 1-{\cal R} \gg 1/\sqrt{N_{0}}$ at times 
$\tau \sim \tau_{h}=1/N_{0}^{2}\to 0$ the annihilation is followed by a sharp 
drop of $h_{s}$, crossing over to the $H$-diffusion-controlled regime, where 
the desorption flux density is governed by the rate of $H$ particles 
diffusion to the system's surface. In the range $\tau_{h} \ll \tau \ll 1$, 
when diffusion of $H$ particles proceeds in a semi-infinite medium 
regime, 
$J\approx h_{0}/\sqrt{\pi\tau}\approx J_{0}/\sqrt{\pi}(\tau_{h}/\tau)^{1/2}$ 
and for surface concentrations I find
$$
h_{s}=(r/\gamma\sqrt{\pi\tau})(1+\varrho + \cdots), \quad 
l_{s}=l_{0}\gamma(1-\varrho + \cdots),
$$
where $\varrho=2r\sqrt{\tau/\pi}$ at $p\ll 1$. At $\tau >1$ the slow 
power-law stage crosses over to the exponential relaxation stage 
$J={\cal A}e^{-\omega_{0}\tau}(1 + e^{-8\omega_{0}\tau}+ \cdots)$ with 
amplitude ${\cal A}=2h_{0}$, and I show that in the limit 
${\cal A}\rightarrow\infty$ the surface concentrations achieve universal 
($\Delta$-{\it independent}) asymptotics
\begin{eqnarray}
h_{s}=\Delta_{c}(1+\phi +e^{-8\omega_{0}\tau}+\cdots), \quad
l_{s}=({\cal A}/\Delta_{c})e^{-\omega_{0}\tau}(1-\phi+\cdots)
\end{eqnarray}
where $|\phi|$ decays as $O(e^{-8\omega_{0}\tau})$ in the range 
$p \leq p_{c}=4/9$ and as $O(e^{-(\pi^{2}/p-\omega_{0})\tau})$ in the range
$p_{c} < p < 1$ [10], and $\Delta_{c}$ obeys the expression [7,9]
$$
\Delta_{c} =\sqrt{\omega_{0}/p}\tan(\sqrt{\omega_{0}p})
$$
($\omega_{0}=\pi^{2}/4$ being the main eigenfrequency of the diffusion 
field relaxation under absorbing boundary conditions). 
This important result implies that in the diffusion-controlled limit 
${\cal A}\rightarrow\infty$ the critical asymptotics 
$h_{s}=h^{c}_{s}=\Delta_{c}$ appears at {\it any} $\Delta$ as a 
{\it precursor} of long-time asymptotics $h_{s}\mid_{\tau\to\infty}=\Delta$, 
and precisely this intermediate asymptotics $h^{c}_{s}$ {\it selects} the 
critical value $\Delta=\Delta_{c}=h^{c}_{s}$ below and above of which the 
relaxation dynamics of $h_{s}$ is qualitatively different: 
at $\Delta < \Delta_{c}$ the surface concentration, $h_{s}$, escaping from 
the asymptotics $h^{c}_{s}$, decreases, reaching its limiting value $\Delta$ 
from {\it above}, whereas at $\Delta > \Delta_{c}$ the surface concentration, 
$h_{s}$, escaping from the asymptotics $h^{c}_{s}$, increases, reaching its 
limiting value $\Delta$ from {\it below} [7]. With neglect of rapidly 
decaying terms of $O(e^{-8\omega_{0}\tau}, e^{(\pi^{2}/p-\omega_{0})\tau})$ 
the exact asymptotic expansion in power of $1/{\cal A}$ has the form [10]
$$
h_{s}=\Delta_{c}(1 + \lambda + \cdots), \quad
l_{s}=({\cal A}/\Delta_{c})e^{-\omega_{0}\tau}(1 - \lambda + \cdots), \quad
N=({\cal A}/\omega_{0})e^{-\omega_{0}\tau}(1 -\mu\lambda + \cdots),
$$
\begin{eqnarray}
J={\cal A}e^{-\omega_{0}\tau}(1-
\sqrt{\omega_{0}}\tanh\sqrt{\omega_{0}}(\Delta-\Delta_{c})^{-1}
\lambda^{2} + \cdots),
\end{eqnarray}
with $\mu=\omega_{0}/\Delta_{c}\approx 1-p$ and $\lambda =
\Delta_{c}(\Delta - \Delta_{c}){\cal A}^{-1}e^{\omega_{0}\tau}.$
These expressions and more detailed analysis in Ref.[7,9] suggest that at 
the crossing of the critical point $\Delta=\Delta_{c}$ not only the sign of 
$\dot{h}_{s}$ changes, but a growth of $h_{s}$ (and with it the whole 
annihilation process) begins to accelerate {\it autocatalytically}, 
$\dot{h}_{s}=\omega_{0}(h_{s}-\Delta_{c})+\cdots$, as a result of 
"switching-on" of the loop of positive feedback: diffusion-induced 
growth of $h_{s}$ accelerates the drop of $l_{s}$, the drop of $l_{s}$ 
accelerates the growth of $h_{s}$ and so on ({\it self-accelerating pairs 
death}). This phenomenon, interpreted in [7] 
as a new type of self-organization, develops the brighter the stronger 
$\Delta$ exceeds $\Delta_{c}$, and in the limit $\Delta \rightarrow \infty$ 
it results in the explosive dynamics - {\it annihilation catastrophe} [8].

{\bf Annihilation catastrophe. Scaling and universality.}
According to (4) in the limit $\Delta \rightarrow \infty$ at the 
self-acceleration stage the rates of growth $\Omega_{Hs}=+d\ln h_{s}/d\tau$ 
and relaxation $\Omega_{Ls}=-d\ln l_{s}/d\tau$ of surface concentrations 
are unambiguously defined by the current value of the reduced pair number 
$n(\tau)=N(\tau)/\Delta$ and are mutually "compensated" so that with an 
accuracy to vanishingly small terms $O(1/n^{2}\Delta)\to 0$ the desorption 
flux relaxation rate $\tau^{-1}_{J}=-d\ln J/d\tau = \Omega_{Ls}-\Omega_{Hs} =
\omega_{0} + \cdots$ is sustained constant. The central result of [8] 
is my finding, that at $\Delta \to \infty$ the relaxation rate 
$\tau^{-1}_{J}$ is sustained constant up to the time moment $\tau_{\star}$ 
of reaching some critical pair number $n_{\star}(p)$, in whose 
vicinity $\delta\tau\propto \Delta^{-2/5} \to 0$ the $\tau^{-1}_{J}$ abruptly 
jumps from $\tau^{-1}_{J}=\omega_{0}$ to $\tau^{-1}_{J} \rightarrow \infty$.
I consider here the two consecutive stages of this phenomenon.

i) {\it Concentration explosion}. Following the work [8], I first show 
that condition $\tau^{-1}_{J}=\omega_{0}$ in the limit 
$\Delta \rightarrow \infty$ leads automatically to the arising of a 
{\it finite-time singularity}. Taking $J={\cal A}e^{-\omega_{0}\tau}$ and 
satisfying (1),(2) to the leading order in $\Delta$, I find
 
\begin{eqnarray}
j=J/\Delta=-\dot{n}=\omega_{0}(1+n), \quad
h_{s}=\Delta_{c}(1+n)/(n-n_{\star}), \quad
l_{s}=\mu\Delta(n-n_{\star}),
\end{eqnarray}

where $n_{\star}=(1-\mu)/\mu \approx p/(1-p)$. It is seen that $h_{s}$
diverges as $n \rightarrow n_{\star}$ and at the point of singularity 
exactly $j_{\star}=\Delta_{c}$. From equality 
$J_{\star}=2h_{0}e^{-\omega_{0}\tau_{\star}}=\Delta\Delta_{c}$
in accord with the numerical data of [10] we obtain

$$
\tau_{\star}=(4/\pi^{2})\ln [2(1+n_{0})/\Delta_{c}].
$$

Taking the origin of time at the point $\tau_{\star}$, i.e. introducing a 
relative time ${\cal T}=\tau-\tau_{\star}$, from (5) we come to remarkably 
simple laws $n=(1/\mu)e^{\omega_{0}|{\cal T}|}-1, 
j=\Delta_{c}e^{\omega_{0}|{\cal T}|}, 
h_{s}=\Delta_{c}/(1-e^{-\omega_{0}|{\cal T}|}), 
l_{s}=\Delta(e^{\omega_{0}|{\cal T}|}-1)$, and, hence, 
$\Omega_{Hs}=\mu(h_{s}-\Delta_{c})=\omega_{0}/(e^{\omega_{0}|{\cal T}|}-1)$, 
whence it follows that for any $p < 1$ at the "distance" from
the singularity point $|{\cal T}| \sim \omega^{-1}_{0}$ the dynamics of
growth of $h_{s}$ changes drastically: the initial exponential increase in
the growth rate 
$\Omega_{Hs}=\omega_{0}e^{-\omega_{0}|{\cal T}|} 
(\Omega_{Ls}\approx\omega_{0})$ turns at $\omega_{0}|{\cal T}| \ll 1$ to a 
singular one
\begin{eqnarray}
\Omega_{Hs}=\Omega_{Ls}=\mu h_{s}=1/|{\cal T}|, \quad
|{\cal T}|\rightarrow 0,
\end{eqnarray}
i.e., {\it concentration explosion} develops in the system.
The striking feature of this phenomenon is that at times 
$\omega_{0}|{\cal T}| \ll 1$ the desorption flux practically reaches its
critical value $J_{\star}$, therefore the concentration explosion develops, 
in fact, at a {\it "frozen"} flux $h_{s}l_{s}=J_{\star}, 
h_{s}=1/\mu|{\cal T}|, l_{s}=\mu J_{\star}|{\cal T}|,$ the consequence of 
which is {\it synchronization} of the growth and relaxation rates 
$\Omega_{Hs}=\Omega_{Ls}=\Omega_{s}$ during the explosion. A singular 
growth of $h_{s}$ at the background of the diffusion flux $J_{\star}$, 
"frozen" on the explosion time scale, gives rise to a singular growth of the 
diffusion "antiflux", the calculation of which yields [8]
$\delta J/J_{\star}\sim - 1/\Delta |{\cal T}|^{3/2}$. As a result, in the 
desorption flux relaxation rate $\tau^{-1}_{J}=\omega_{0}+[\tau^{-1}_{J}]$ 
the singularly growing term arises

\begin{eqnarray}
[\tau^{-1}_{J}]\sim 1/\Delta|{\cal T}|^{5/2},
\end{eqnarray}
whence it follows that the scenario described above holds down to
$|{\cal T}|_{cat}\sim \Delta^{-2/5} \to 0$, beyond which a catastrophic 
growth of $\tau^{-1}_{J}$ begins.

ii) {\it Two universality classes of passing through the point of 
singularity}. Remarkably, the dynamics of passing through the point of
singularity can be described analytically (which, in itself, is a rather rare 
event [11]) and it appears quite beautiful. I show analytically and confirm 
numerically that at the singularity point ${\cal T}=0$ the rate of explosion 
$\Omega_{Hs}$ reaches a maximum $\Omega^{M}_{Hs}$, the dynamics of passing 
through which proceeds in one of the two characteristic scaling regimes, 
depending on the value of parameter ${\cal K}=p^{3/2}\Delta/\Delta_{c}$.

{\it 1. Annihilation catastrophe at "frozen" flux (${\cal K} \gg 1$)}. 
In the limit ${\cal K} \gg 1$ at times $|{\cal T}| \ll {\cal T}_{L}=p$ 
the particle number $n=n_{\star}(1+\omega_{0}|{\cal T}|/p +\cdots)$ is, 
in fact, "frozen" $n\approx n_{\star}$ and so in the process of explosion 
in the contracting as $\delta x_{L} \sim \sqrt{|{\cal T}|/p}$ nearsurface 
layer there arises a sharp gradient of $L$ particles. I show that in this 
limit up to ${\cal T} \ll p$ the flux and the particle number are retained 
"frozen", $h_{s}l_{s}=J_{\star}$ and, hence, the explosion develops 
synchronouosly, $\Omega_{Hs}=\Omega_{Ls}=\Omega_{s}$, both {\it before} and 
{\it after} the singularity point. Equalizing the rates of change of $H$ and 
$L$ particles diffusion fluxes leads to the scaling 
\begin{eqnarray}
\Omega_{s}=\Omega^{M}_{s}S({\cal T}/{\cal T}_{f}), \quad 
h_{s}/h^{M}_{s}=l^{M}_{s}/l_{s}=w({\cal T}/{\cal T}_{f}), 
\end{eqnarray}
where the scaling functions $S(\zeta)=1/\sqrt{1+\zeta^{2}}$ and 
$w(\zeta)=\zeta+\sqrt{1+\zeta^{2}}$, the characteristic time 
${\cal T}_{f}=1/\Omega^{M}_{s}$, and the amplitudes 
$\Omega^{M}_{s}, h^{M}_{s}$ and $l^{M}_{s}$ at the explosion maximum 
$$
h^{M}_{s}=(2/\mu)\Omega^{M}_{s}=p^{-1/4}\sqrt{\Delta\Delta_{c}}, \quad
l^{M}_{s}=p^{1/4}\sqrt{\Delta\Delta_{c}}.
$$
Two striking features of this result are symmetrical universalization 
of $\Omega_{s}=1/|{\cal T}|$ beyond the scope of interval 
$[-{\cal T}_{f},{\cal T}_{f}], {\cal T}_{f}\to 0$ and remarkable symmetry 
${\cal T}\leftrightarrow -{\cal T}, w\leftrightarrow 1/w$. 

Singular "flash" of $\Omega_{s}$ leads to a singular jump in $\tau^{-1}_{J}$, 
the dynamics of which is described by the scaling 
\begin{eqnarray}
[\tau^{-1}_{J}]= [\tau^{-1}_{J}]_{\star}W({\cal T}/{\cal T}_{f}), \quad
[\tau^{-1}_{J}]_{\star}=c_{1}(\Omega^{M}_{s})^{5/2}/\Delta \propto
\mu^{5/4}p^{-5/8}\Delta^{1/4}, 
\end{eqnarray}
where in accord with the scaling function $W(\zeta)= 
c_{2}\int_{0}^{\infty}d\theta/\sqrt{\theta}[1+(\zeta-\theta)^{2}]^{3/2}$ 
the value of $[\tau^{-1}_{J}]$ reaches at ${\cal T}\sim {\cal T}_{f}$ 
a maximum ${\rm max}[\tau^{-1}_{J}]\sim [\tau^{-1}_{J}]_{\star}
\propto \Delta^{1/4}\to \infty$, and then it decays as 
$\sim 1/\sqrt{p{\cal T}}$. Essentially, that 
${\rm max}[\tau^{-1}_{J}]\propto p^{-5/8}$, so the less is $p$ the more 
brightly the effect is displayed. 

{\it 2.Annihilation catastrophe at uniform $L$ particles distribution. 
Flux breaking effect (${\cal K} \ll 1$).} If $\Delta\to\infty$ but $p\to 0$ 
so that ${\cal K} \ll 1$, the distribution of $L$ species is retained uniform 
$(l_{s}=N, \Omega_{Ls}=h_{s})$ and the final stage of explosion is changed 
radically. I demonstrate that in this case an intermediate scaling takes 
place
\begin{eqnarray}
\Omega_{Hs}=\Omega^{M}_{Hs}Q({\cal T}/{\cal T}_{*}), \quad
h_{s}=h^{M}_{s}v_{H}({\cal T}/{\cal T}_{*}), \quad
l_{s}=l^{M}_{s}v_{L}({\cal T}/{\cal T}_{*})
\end{eqnarray}
where the characteristic time ${\cal T}_{*}=1/\Omega^{M}_{Hs}$ and amplitudes 
$\Omega^{M}_{Hs}, h^{M}_{s}$ and $l^{M}_{s}$ at the explosion maximum 
$$
h^{M}_{s}=q\Omega^{M}_{Hs}=b_{H}\Delta^{2/3}, \quad
l^{M}_{s}=b_{L}\Delta^{1/3},
$$
with $q=2.24, b_{H}=1.63$ and $b_{L}=0.73$. It is seen that in the limit 
${\cal K} \ll 1$ the singular "flash" of $\Omega_{Hs}$ occurs much sharper, 
leading to more abrupt jump in $\tau^{-1}_{J}$ 
\begin{eqnarray}
[\tau^{-1}_{J}]=[\tau^{-1}_{J}]_{\star}V({\cal T}/{\cal T}_{*}), \quad
[\tau^{-1}_{J}]_{\star}=(q-1)\Omega^{M}_{Hs}\propto \Delta^{2/3},
\end{eqnarray}
which, in turn, gives rise to a qualitatively new phenomenon: 
the finite jump of the flux by a factor of $\approx 2$ for the time 
$|{\cal T}|_{break}\sim{\cal T}_{*}\propto \Delta^{-2/3}\to 0$ 
({\it flux breaking effect}).

As in the limit ${\cal K} \gg 1$ at ${\cal T} \gg p$ the relaxation rate 
$\tau^{-1}_{J}$ reaches its $L$-diffusion-controlled limit 
$\tau^{-1}_{J}\sim \omega_{0}/p$, and in the limit ${\cal K}\ll 1$ it 
always grows, I point out finally (and demonstrate numerically) that 
independently of ${\cal K}$ in the limit 
$p\rightarrow 0, \Delta\rightarrow\infty$ there arises a most spectacular 
consequence of the annihilation catastrophe:{\it an abrupt, practically 
instantaneous} (on the scale of $\omega_{0}$) 
{\it disappearance of the flux}.    

In conclusion, I believe that annihilation catastrophe may pretend to be 
not only one of the most dramatic manifestations of the reaction-diffusion 
interplay, but it is a {\it new type of catastrophe} wherein two explosive 
processes ($\Omega_{Hs}$ and $\Omega_{Ls}$) are developing simultaneously, 
effectively "compensating" one another so that for an external observer 
($J$) the explosion dynamics goes unnoticed up to the point of singularity, 
in the vicinity of which decompensation of explosions is manifested as a 
sudden singular jump.
\\
\\
This work was supported by the RFBR Grant No. 02-03-33122-a.
\\
\\ 
{\bf References}: 
[1] For a review see E. Kotomin and V. Kuzovkov,
{\it Modern Aspects of Diffusion Controlled Reactions: Cooperative
Phenomena in Bimolecular Processes} (Elsevier, Amsterdam, 1996); 
[2] A.A. Ovchinnikov and Ya.B. Zeldovich, Chem. Phys. 
{\bf 28}, 215 (1978); [3] D. Toussaint and F. Wilczek, J. Chem. Phys. 
{\bf 78}, 2642 (1983); [4] L. Galfi and Z. Racz, Phys. Rev. A {\bf 38}, 3151 
(1988); [5] E. Ben-Naim and S. Redner, J. Phys. A {\bf 25}, L575 (1992); [6] 
S. Cornell and M. Droz, Phys. Rev. Lett. {\bf 70}, 3824 (1993); [7] 
B.M. Shipilevsky, Phys. Rev. Lett. {\bf 73}, 201 (1994); 
[8] B.M. Shipilevsky, Phys. Rev. Lett. {\bf 82}, 4348 (1999);
[9] B.M. Shipilevsky, J. Phys. A {\bf 30}, L471 (1997); [10] B.M. Shipilevsky 
(to be published); [11] L.P. Kadanoff, {\it Singularities and Blowups}, 
Phys. Today, 11 (1997).
\end{document}